\documentclass[nofootinbib,aps,twocolumn,showpacs,amsmath,amssymb,unsortedaddress]{revtex4-1}
\usepackage{amssymb}
\usepackage{amsmath}
\usepackage{epsfig}
\usepackage{amsfonts}
\usepackage{color}
\usepackage{float}
\usepackage{latexsym}
\usepackage{mathrsfs}
\usepackage{feynmf}
\usepackage{balance}

\begin{document}
\title{ Factorization law for two lower bounds of concurrence  }
\author{Sayyed Yahya Mirafzali}
\email{yahya.mirafzali@stu-mail.um.ac.ir} 
\author {Iman Sargolzahi}
\affiliation{Department of Physics, Ferdowsi University of Mashhad,  Mashhad, Iran}

\author{Ali Ahanj}
\email{ahanj@ipm.ir} 
\affiliation{Khayyam Institute of Higher Education,  Mashhad, Iran}
\affiliation{School of Physics, Institute for Research in Fundamental Science(IPM), P. O. Box 19395-5531, Tehran, Iran.}

\author{Kurosh Javidan}
%\affiliation{Department of Physics, Ferdowsi University of Mashhad,  Mashhad, Iran}
\author{Mohsen Sarbishaei}
\affiliation{Department of Physics, Ferdowsi University of Mashhad,  Mashhad, Iran}

%%%%%%%%%%%%%%%%%%%%%%%%%%%%%%%%%%%%%%%%%%%%%%%%%%%%%%%%%%%%%%%%%%%%%%%%%%%%%%%%%%%%%%%%
\begin{abstract}

We study the dynamics of two lower bounds of concurrence in bipartite quantum systems when one party goes 
through an arbitrary channel. We show that these lower bounds obey the factorization law similar to that 
of [Konrad \textit{et al.}, Nat. Phys. \textbf{4}, 99 (2008)].  We also, discuss the application 
of this property, in an example.
  
\end{abstract}

%%%%%%%%%%%%%%%%%%%%%%%%%%%%%%%%%%%%%%%%%%%%%%%%%%%%%%%%%%%%%%%%%%%%%%%%%%%%%%%%%%%%

\pacs{03.65.Ud, 03.67.Mn}
\maketitle

\section{Introduction}

Entanglement, one of the important features of quantum systems which does not exist classically, has been known as a key 
resource for some quantum computation and information processes. But the entanglement of a system 
changes due to its unavoidable interactions with environment. To study the entanglement changes, one needs to make   
use of an entanglement measure in order to specify the entanglement amount of a system. Unfortunately, most of 
the measures having been proposed for quantification of entanglement can not be computed in general, and  because of this, many lower and upper bounds, which can be computed easily, have been introduced for these entanglement measures. Using these bounds, one can estimate the amount of entanglement. 

In Ref.~\cite{1} Konrad \textit{et al.} have provided
a factorization law for concurrence which is one of the remarkable entanglement measures. They have shown that the 
concurrence of a two-qubit state, when one of its qubits goes through an arbitrary quantum channel, is equal to the product of its initial concurrence and concurrence of the maximally entangled state undergoing the effect of the same quantum channel. 
Then Li \textit{et al.}~\cite{2} have shown that the generalization of the above factorization law to arbitrary dimensional bipartite states only leads to an upper bound for the concurrence of the system. If, beside this upper bound we have a lower bound obeying a similar factorization law, then we can make better use of this useful dynamical property. So it will be valuable to seek for such entanglement lower bounds.

In section II, we introduce the concurrence and some of its lower bounds. Next, in section III, we briefly review the results of Ref.~\cite{1,2}. Then, in sections IV and V, we investigate the factorization property of the lower bounds introduced in section II. In section VI, as an application, we discuss an example.  Finally, we give some conclusions in section VII.

%%%%%%%%%%%%%%%%%%%%%%%%%%%%%%%%%%%%%%%%%%%%%%%%%%%%%%%%%%%%%%%%%%%%%%%%%%%%%%%%%%%%%%%%%%%%%%%

\section{concurrence and some of its lower bounds}

For a pure bipartite state $ \vert\Psi\rangle $, $ \vert\Psi\rangle\in\mathcal{H}_{A}\otimes \mathcal{H}_{B}\ $,
concurrence is defined as~\cite{4}:
\begin{equation}
C\left( \Psi\right)=\sqrt{2 [\langle\Psi\vert\Psi\rangle^{2} -tr\rho_{r}^{2} ]}\,,
\label{1}
\end{equation}
where $ \rho_{r} $ is the reduced density operator obtained by tracing over either   
subsystems A or B.  
Concurrence of $\vert\Psi\rangle$ can also be written in terms of the expectation value of an 
observable with respect to two identical copies of $ \vert\Psi\rangle $~\cite{4,5,6}: 
\begin{align}
C\left( \Psi\right)=\sqrt{_{AB}\langle\Psi\vert_{A'B'}\langle\Psi\vert{\cal A} \vert\Psi\rangle_{AB}\vert\Psi\rangle_{A'B'}} \,,\notag \\
{\cal A}=4 P_{-}^{AA'}\otimes P_{-}^{BB'}\,,\qquad\qquad
\label{2}        
\end{align}
where $ P_{-}^{AA'} $ ($ P_{-}^{BB'} $) is the projector onto the antisymmetric subspace of
 $ \mathcal{H}_{A}\otimes \mathcal{H}_{A'}\ $ ($ \mathcal{H}_{B}\otimes \mathcal{H}_{B'}\ $).
A possible decomposition of $ {\cal A} $ is 
\begin{align}
{\cal A}=\sum_{i<j,m<n}\vert\chi_{ij,mn}\rangle\langle\chi_{ij,mn}\vert \,,\qquad\qquad\qquad\notag \\
\vert\chi_{ij,mn}\rangle\ =\large(\vert ij\rangle - \vert ji\rangle)_{AA'}\large(\vert mn\rangle - \vert nm\rangle)_{BB'}\,,
\label{3}
\end{align}
where $ \vert i\rangle $ and $ \vert j\rangle $ ( $  \vert m\rangle $ and $ \vert n\rangle $ )are
two different members of an orthonormal basis of the A (B) subsystem (instead of the index $\alpha$ in reference~\cite{4}, we use the indexes $ij,mn$ because it seams most convenient for the future usage).

For mixed states, the concurrence is defined as follows ~\cite{4}:
\begin{align}
C(\rho)= \min_{\lbrace p_{k},\Psi_{k}\rbrace} \sum_{k}p_{k}C( \Psi_k )\,, \qquad\qquad\notag \\
\rho =\sum_{k}p_{k} \vert\Psi_{k}\rangle\langle\Psi_{k}\vert\,, \qquad p_{k}\geq 0\,, \qquad\sum_{k}p_{k} =1\,,
\label{4}
\end{align}
where the minimum is taken over all decompositions of $ \rho $ into pure states $ \vert\Psi_{k}\rangle$. Like most of  
the other entanglement measures, $C(\rho)$ can not be computed in general, i.e., in general, one can not find the optimal decomposition of $ \rho $ minimizing Eq. (\ref{4}). Any numerical effort for finding the optimal decomposition, is equivalent to find an upper bound for $C(\rho)$. So, some lower bounds have been introduced for $C(\rho)$ (e.g.~\cite{7,new}).

It has been shown that  
\begin{align}
ALB_{ij,mn}(\rho)\equiv \min_{\lbrace p_{k},\vert\Psi_{k}\rangle\rbrace} \sum_{k}p_{k}\vert\langle\chi_{ij,mn} \vert\Psi_{k}\rangle\vert\Psi_{k}\rangle\vert ,
\label{5}
\end{align}
is a lower bound of concurrence ($ALB$ is the abbreviation of the \textit{Algebraic Lower Bound})~\cite{4,7,8}. $ALB_{ij,mn}(\rho)$ can be computed analytically; $ ALB_{ij,mn}(\rho)=\max \lbrace 0, \mathscr{S}_{1}^{ij,mn}-\sum_{l>1}\mathscr{S}_{l}^{ij,mn}\rbrace\ $~\cite{4}. $ \mathscr{S}_{l}^{ij,mn} $ are the singular values of matrix $ T^{ij,mn} $ in decreasing order. $ T^{ij,mn} $'s entries are defined as $\qquad T_{rs}^{ij,mn}\equiv\sqrt{\lambda_{r}\lambda_{s}}\langle\chi_{ij,mn}\vert\Phi_{r}\rangle\vert\Phi_{s}\rangle\ $ , where  $ \vert\Phi_{r}\rangle $ and $ \lambda_{r} $ are eigenvectors and eigenvalues of $ \rho $, respectively.

Another lower bounds of concurrence are those introduced in reference~\cite{8}. In this reference, it has been shown that in terms of  two identical copies of an arbitrary mixed state $ \rho_{AB} $ we have     
\begin{align}
C^{2}(\rho_{AB})\geq MLB^{2}_{(k)ij,mn}(\rho)\equiv tr\left( \rho_{AB}\otimes \rho_{A'B'} V_{(k)ij,mn}\right),\notag \\
 k=1,2   ,\quad\quad\qquad\qquad\quad\quad\qquad\qquad\notag \\
V_{(1)ij,mn} =4 P_{-ij}^{AA'}\otimes\left(P_{-mn}^{BB'}-P_{+mn}^{BB'}\right)\,,\quad\quad\qquad\notag \\
V_{(2)ij,mn} =4 \left(P_{-ij}^{AA'}-P_{+ij}^{AA'}\right) \otimes P_{-mn}^{BB'}\,.\quad\quad\qquad\notag \\
\label{6}
\end{align}
($MLB$ is the abbreviation of the \textit{Measurable lower bound}) where $ 2P_{-ij}^{AA'}=\large(\vert ij\rangle - \vert ji\rangle)\large(\langle ij\vert -\langle ji\vert) $ and $ 2P_{+ij}^{AA'}=\large(\vert ij\rangle + \vert ji\rangle)\large(\langle ij\vert +\langle ji\vert)+2\vert ii\rangle\langle ii\vert+2\vert jj\rangle\langle jj\vert $ operate on $\mathcal{H}_{A}\otimes \mathcal{H}_{A'}$ whereas $ 2P_{-mn}^{BB'}=\large(\vert mn\rangle - \vert nm\rangle)\large(\langle mn\vert -\langle nm\vert) $ and $ 2P_{+mn}^{BB'}=\large(\vert mn\rangle + \vert nm\rangle)\large(\langle mn\vert +\langle nm\vert)+2\vert mm\rangle\langle mm\vert+2\vert nn\rangle\langle nn\vert $  operate on $\mathcal{H}_{B}\otimes \mathcal{H}_{B'}$ ($\vert i\rangle, \vert j\rangle, \vert m\rangle$ and $\vert n\rangle$ were introduced in Eq. (\ref{3})). The above expression means that measuring $ V_{(k)ij,mn} $ on two identical copies of $ \rho $, i.e.  
$ \rho\otimes\rho $, gives us a measurable lower bound on $ C^{2}(\rho) $.

In reference~\cite{9}, another lower bound of concurrence was introduced. There, it was shown that:
\begin{align}
\tau(\rho) \equiv\sum_{i<j,m<n} C_{ij,mn}^{2}(\rho)\leq C^{2}(\rho)\,,\qquad\quad\notag \\
 C_{ij,mn}(\rho) =\min_{\left\lbrace p_{k},\vert\psi_{k}\rangle\right\rbrace }
\sum_{k} p_{k}\vert\langle\Psi_{k}\vert L_{A,ij}\otimes L_{B,mn}\vert\Psi_{k}^{\ast}\rangle\vert ,
\label{7} 
\end{align}
where $ L_{A,ij} $ and $ L_{B,mn} $ are the generators of $ SO(d_{A}) $ and $ SO(d_{B})$ respectively 
($d_{A}$($d_{B}$) is the dimension of $\mathcal{H}_{A}$($\mathcal{H}_{B}$)), and $ \vert\Psi_{k}^{\ast}\rangle $ is the complex conjugate of
$ \vert\Psi_{k}\rangle $ in the computational basis. In this basis $ L_{A,ij} $ and $ L_{B,mn} $ are~\cite{10}:
\begin{align}
L_{A,ij} &= \vert i\rangle_{A}\langle j\vert -\vert j\rangle_{A}\langle i\vert\,,\notag \\
L_{B,mn} &= \vert m\rangle_{B}\langle n\vert -\vert n\rangle_{B}\langle m\vert\,.
\end{align}

%%%%%%%%%%%%%%%%%%%%%%%%%%%%%%%%%%%%%%%%%%%%%%%%%%%%%%%%%%%%%%%%%%%%%%%%%%%%%%%%%%%%%%

\section{factorization of the concurrence}
According to the Schmidt decomposition, any pure bipartite state $ \vert\Psi\rangle $, $ \vert\Psi\rangle\in\mathcal{H}_{A}\otimes \mathcal{H}_{B}$, 
can be expressed as 
\begin{align}
\vert\Psi\rangle=\sum_{i=1}^{d}\sqrt{\omega_{i}}\vert\alpha_{i}\beta_{i}\rangle\,, \quad
0\leq\sqrt{\omega_{i}}\leq 1\,, \quad \sum_{i=1}^{d}\omega_{i} =1,\,
\label{9}
\end{align}
where $d=min(d_{A}, d_{B})$.

We can rewrite this $ \vert\Psi\rangle$ as $ \vert\Psi\rangle=(M\otimes I)\vert\phi^{+}\rangle$ where $\vert\phi^{+}\rangle=\sum_{i=1}^{d}\frac{1}{\sqrt{d}}\vert\alpha_{i}\beta_{i}\rangle$ is a maximally entangled state and $M=\sqrt{d}\sum_{i=1}^{d}\sqrt{\omega_{i}}\vert\alpha_{i}\rangle\langle\alpha_{i}\vert$. 

Assume that the second part of this state goes through an arbitrary channel $\mathcal{S}$, then this state transforms to $\rho'=\frac{({\bf{1}}\otimes\mathcal
{S})|\Psi\rangle\langle\Psi|}{p'}$ where $p'=tr[({\bf{1}}\otimes\mathcal
{S})|\Psi\rangle\langle\Psi|]$. Since $M$ and $\mathcal{S}$ act on two different parts of $\vert\Psi\rangle$, $\rho'$ can be written as
$\rho'=\frac{(M\otimes{\bf{I}})\rho_{\mathcal{S}}(M^{\dagger}\otimes{\bf{I}})}{p}$ where $\rho_{\mathcal{S}}=\frac{({\bf{1}}\otimes\mathcal{S})|\phi^{+}\rangle\langle\phi^{+}|}{p''}$, 
$p=tr[(M\otimes{\bf{I}})\rho_{\mathcal{S}}(M^{\dagger}\otimes{\bf{I}})]$, $p''=tr[
({\bf{1}}\otimes\mathcal{S})|\phi^{+}\rangle\langle\phi^{+}|]$ and $p'=pp''$. 

By using these relations, for any two-qubit state $\vert\Psi\rangle$, Konrad \textit{et al.}~\cite{1} have proved the following factorization law~\cite{11}:
\begin{align}
C[({\bf{1}}\otimes\mathcal
{S})|\Psi\rangle\langle\Psi|]=C[({\bf{1}}\otimes\mathcal
{S})|\phi^{+}\rangle\langle\phi^{+}|]C(\Psi).
\label{10}
\end{align}    

The right hand side of the above equation is factorized into two independent parts. The first part is the concurrence of $\vert\phi^{+}\rangle$ after going through the channel $({\bf{1}}\otimes\mathcal{S})$, which is independent of the initial state $\vert\Psi\rangle$, and the second part is the concurrence of the initial state $\vert\Psi\rangle$(before going into the channel). So, if we know the concurrence of $\vert\phi^{+}\rangle$, after one of its qubits goes through a channel $\mathcal
{S}$, we know, up to the factor $C(\Psi)$, the concurrence of any arbitrary state $\vert\Psi\rangle$ undergoing true the same quantum channel.

For higher dimensional bipartite systems, Lie \textit{et al.}~\cite{2} have shown that the above equality changes to the following inequality
\begin{align}
C[({\bf{1}}\otimes\mathcal
{S})|\Psi\rangle\langle\Psi|]\leq\frac{d_{B}}{2}
C[({\bf{1}}\otimes\mathcal{S})|\phi^{+}\rangle\langle\phi^{+}|]C(\Psi).
\label{11}
\end{align} 

For the $d_{A}\times 2$ dimensional states, we have the equality instead of the inequality in the above relation.  But, in general, the concurrence of $({\bf{1}}\otimes\mathcal{S})|\phi^{+}\rangle\langle\phi^{+}|$ provides only an upper bound for $C[({\bf{1}}\otimes\mathcal
{S})|\Psi\rangle\langle\Psi|]$. We point out that in relations (\ref{10}) and (\ref{11}), instead of $\vert\phi^{+}\rangle$, we can use any other maximally entangled state.  

It is also interesting to investigate a similar relations for the lower bounds of concurrence. In the next section, we study the factorization property of the lower bounds introduced in the previous section.

%%%%%%%%%%%%%%%%%%%%%%%%%%%%%%%%%%%%%%%%%%%%%%%%%%%%%%%%%%%%%%%%%%%%%%%%%%%%%%%%%%%%%%%

\section{factorization of the lower bounds of concurrence}
Let us at first consider the lower bound introduced in expression (\ref{6}). From this relation we have
\begin{widetext} 
\begin{align}
p^{2}MLB^{2}_{(1)ij,mn}(\rho')&= p^{2}tr\left( \rho'_{AB}\otimes\rho'_{A'B'} V_{(1)ij,mn}\right)\notag \\
&=tr\big[(M_{A}\otimes{\bf{I}}_{B})\rho_{\mathcal{S}AB}(M^{\dagger}_{A}\otimes{\bf{I}}_{B})
\otimes(M_{A'}\otimes{\bf{I}}_{B'})\rho_{\mathcal{S}A'B'}(M^{\dagger}_{A'}\otimes{\bf{I}}_{B'})V_{(1)ij,mn}\big] \notag \\
&=tr\big[(M_{A}\otimes{\bf{I}}_{B}\otimes M_{A'}\otimes{\bf{I}}_{B'})(\rho_{\mathcal{S}AB}\otimes\rho_{\mathcal{S}A'B'})(M^{\dagger}_{A}\otimes{\bf{I}}_{B}\otimes M^{\dagger}_{A'}\otimes{\bf{I}}_{B'})V_{(1)ij,mn}\big]\notag \\
&=tr\big[(\rho_{\mathcal{S}AB}\otimes\rho_{\mathcal{S}A'B'})(M^{\dagger}_{A}\otimes{\bf{I}}_{B}\otimes M^{\dagger}_{A'}\otimes{\bf{I}}_{B'})
V_{(1)ij,mn}(M_{A}\otimes{\bf{I}}_{B}\otimes M_{A'}\otimes{\bf{I}}_{B'})\big] \notag \\
&=d^{2}\omega_{i}\omega_{j}tr\big[\rho_{\mathcal{S}AB}\otimes\rho_{\mathcal{S}A'B'}V_{(1)ij,mn}\big].
\label{14} 
\end{align}
\end{widetext}
In order to obtain the last equality, we have used $(M_{A}^{\dagger}\otimes M_{A'}^{\dagger})P_{-ij}^{AA'}(M_{A}\otimes M_{A'})=d^{2}\omega_{i}\omega_{j}P_{-ij}^{AA'}$ where $P_{-ij}^{AA'}$ is written in the Schmidt basis, i.e. we choose 
$\vert i\rangle=\vert\alpha_{i}\rangle$ and $\vert j\rangle=\vert\alpha_{j}\rangle$ in construction of $P_{-ij}^{AA'}$. Also, writing $P_{-mn}^{BB'}$ and $P_{+mn}^{BB'}$ in the Schmidt basis, we have $MLB^{2}_{(1)ij,mn}(\vert\Psi\rangle)=4\omega_{i}\omega_{j}\delta_{im}\delta_{jn}$. Using this relation, Eq. (\ref{14}) can be written in the form:
\begin{align}
MLB^{2}_{(1)ij,mn}(({\bf{1}}\otimes\mathcal
{S})|\Psi\rangle\langle\Psi|)\,\quad\qquad\qquad\qquad\qquad\qquad\notag \\
=\frac{d^{2}}{4}MLB^{2}_{(1)ij,mn}(({\bf{1}}\otimes\mathcal{S})|\phi^{+}\rangle\langle\phi^{+}|)
 MLB_{(1)ij,ij}^{2}(\vert\Psi\rangle).
\label{15} 
\end{align}

The above equation(which is our main result) is similar to Eq. (\ref{10}), so $MLB^{2}_{(1)ij,mn}(\rho)$ have the same factorization property as concurrence, i.e.  knowing the effect of $({\bf{1}}\otimes\mathcal{S})$ on the $MLB^{2}_{(1)ij,mn}(\rho)$ when the initial state is $\vert\Phi^{+}\rangle$, we know this effect for any other initial state $\vert\Psi\rangle$, up to a factor $MLB_{(1)ij,ij}^{2}(\vert\Psi\rangle)$.

For the $MLB^{2}_{(2)ij,mn}(\rho)$, we obtain exactly the same result as above if instead of the second part, the first
part of the state $\vert\Psi\rangle$ goes through the channel $\mathcal{S}$. 

Now we discuss the factorization property of $ALB_{ij,mn}(\rho')$. We use a similar method as Ref.~\cite{2}, namely, at first we restrict ourselves to those cases where $\rho_{\mathcal{S}}$ is a pure state i.e. $\rho_{\mathcal{S}}=\vert\psi\rangle\langle\psi\vert$. In this cases $\rho'$ is also a pure state i.e. 
$\rho'\equiv\vert\psi'\rangle\langle\psi'\vert=\frac{(M\otimes{\bf{I}})\vert\psi\rangle\langle\psi\vert(M^{\dagger}\otimes{\bf{I}})}{p}$. From Eq. (\ref{5}) we have   
\begin{align}
p ALB_{ij,mn}(\vert\psi'\rangle)&=p\vert\langle\chi_{ij,mn} \vert\psi'\rangle\vert\psi'\rangle\vert \,\notag \\
&=\vert\langle\chi_{ij,mn} \vert M\otimes{\bf{I}}\otimes M\otimes{\bf{I}}\vert\psi\rangle\vert\psi\rangle\vert\,\notag \\
&=d\sqrt{\omega_{i}\omega_{j}} ALB_{ij,mn}(\rho_{\mathcal{S}}),
%=\frac{d}{2}ALB_{\alpha}(\vert\Psi\rangle) ALB_{\alpha}(\rho_{\mathcal{S}})\,\qquad\qquad\quad
\label{161}
\end{align}
where we used $(M^{\dagger}\otimes{\bf{I}}\otimes M^{\dagger}\otimes{\bf{I}})\vert\chi_{ij,mn}\rangle\langle\chi_{ij,mn}\vert 
(M\otimes{\bf{I}}\otimes M\otimes{\bf{I}})=d^{2}\omega_{i}\omega_{j}\vert\chi_{ij,mn}\rangle\langle\chi_{ij,mn}\vert$ and  
$\vert\chi_{ij,mn}\rangle$ is written in the Schmidt basis. Using $ALB_{ij,mn}^{2}(\vert\Psi\rangle)=4\omega_{i}\omega_{j}\delta_{im}\delta_{jn}$, we obtain 
\begin{align}
p ALB_{ij,mn}(\vert\psi'\rangle)=\frac{d}{2}ALB_{ij,ij}(\vert\Psi\rangle) ALB_{ij,mn}(\rho_{\mathcal{S}}).\,
\label{16}
\end{align}

Next, we consider the general case where $\rho_{\mathcal{S}}$ is a mixed state. Corresponding to any pure state decomposition of $\rho_{\mathcal{S}}$ as $\rho_{\mathcal{S}}=\sum_{k}p_{k}\vert\psi_{k}\rangle\langle\psi_{k}\vert$, there exist a pure state decomposition for $\rho'$ in terms of pure states $\vert\psi'_{k}\rangle=\frac{(M\otimes{\bf{I}})\vert\psi_{k}\rangle}{\sqrt{pq_{k}}}$, $q_{k}=tr(\frac{(M\otimes{\bf{I}})\vert\psi_{k}\rangle\langle\psi_{k}\vert (M^{\dagger}\otimes{\bf{I}})}{p})$, such that $\rho'=\sum_{k}p_{k}q_{k}\vert\psi_{k}'\rangle\langle\psi_{k}'\vert$. 
Thus, by using the same arguments as before, for any $\vert\psi_{k}'\rangle$, we have $pq_{k} \vert\langle\chi_{ij,mn} \vert\psi_{k}'\rangle\vert\psi_{k}'\rangle\vert=d\sqrt{\omega_{i}\omega_{j}}\vert\langle\chi_{ij,mn} \vert\psi_{k}\rangle\vert\psi_{k}\rangle\vert$. Now, assume $\rho_{\mathcal{S}}=\sum_{k}p_{k}\vert\psi_{k}\rangle\langle\psi_{k}\vert$ is 
the optimal pure state decomposition which gives $ALB_{ij,mn}(\rho_{\mathcal{S}})$, i.e. 
$ALB_{ij,mn}(\rho_{\mathcal{S}})=\sum_{k}p_{k}\vert\langle\chi_{ij,mn} \vert\psi_{k}\rangle\vert\psi_{k}\rangle\vert$ so
$p \sum_{k}p_{k}q_{k}\vert\langle\chi_{ij,mn} \vert\psi_{k}'\rangle\vert\psi_{k}'\rangle\vert=d\sqrt{\omega_{i}\omega_{j}}ALB_{ij,mn}(\rho_{\mathcal{S}})$. 
But $\sum_{k}p_{k}\vert\psi_{k}'\rangle\langle\psi_{k}'\vert$ is \textit{not} necessarily the optimal pure state decomposition of $\rho'$ such that $ALB_{ij,mn}(\rho')=\sum_{k}p_{k}\vert\psi_{k}'\rangle\langle\psi_{k}'\vert$. Therefore in general       
\begin{align}
ALB_{ij,mn}(({\bf{1}}\otimes\mathcal
{S})|\Psi\rangle\langle\Psi|)\,\qquad\qquad\qquad\qquad\qquad\quad\notag \\
\leq \frac{d}{2}ALB_{ij,ij}(\vert\Psi\rangle) ALB_{ij,mn}(({\bf{1}}\otimes\mathcal{S})|\phi^{+}\rangle\langle\phi^{+}|).\,\quad
\label{17}
\end{align}

In the cases where $M^{-1}$ exists, i.e. when in Eq. (\ref{9}) for all $\omega_{i}$ we have $\omega_{i}\neq 0$, as for the $d_{A}\times 2$ dimensional systems (the case of the separable initial states is not of interest), corresponding to any pure state decomposition for $\rho'$, there is a pure state decomposition for $\rho_{\mathcal{S}}$ and vice versa, namely, for any $\vert\psi'_{k}\rangle$ in the experssion $\rho'=\sum_{k}p_{k}\vert\psi_{k}'\rangle\langle\psi_{k}'\vert$ we have $\vert\psi_{k}\rangle=\sqrt{p}(M^{-1}\otimes{\bf{I}})\vert\psi'_{k}\rangle$ such that $\rho_{\mathcal{S}}=\sum_{k}p_{k}\vert\psi_{k}\rangle\langle\psi_{k}\vert$. So, if the $\rho_{\mathcal{S}}=\sum_{k}p_{k}\vert\psi_{k}\rangle\langle\psi_{k}\vert$ is the optimal decomposition  for $ALB_{ij,mn}(\rho_{\mathcal{S}})$ then $\sum_{k}p_{k}\vert\psi_{k}'\rangle\langle\psi_{k}'\vert$ is the optimal pure state decomposition of $\rho'$ for $ALB_{ij,mn}(\rho')$. Therefore, in Eq. (\ref{17}) we have an equality instead of the inequality.
 
%%%%%%%%%%%%%%%%%%%%%%%%%%%%%%%%%%%%%%%%%%%%%%%%%%%
\section{factorization of the lower bound of squared concurrence ($\tau$)}
In Ref.~\cite{3} Liu \textit{et al.} have shown that $\tau$ (Eq. (\ref{7})), for a $d\times d$ bipartite quantum state, obeys the relation  
\begin{equation}
\tau(({\bf{1}}\otimes\mathcal
{S})|\Psi\rangle\langle\Psi|)\leq\frac{d^{2}}{4}\tau(({\bf{1}}\otimes\mathcal
{S})|\phi^{+}\rangle\langle\phi^{+}|)C^{2}(\Psi).
\label{12}
\end{equation}
The above relation is the factorization law for $\tau$ similar to the Eq. (\ref{11}) which is for the concurrence itself. 

Now, we show that $ALB_{ij,mn}(\rho)$ is closely related to $\tau$; For an arbitrary $ \vert\Psi\rangle $, according to the definition of $\vert\chi_{ij,mn}\rangle$ in 
Eq. (\ref{3}), it can be seen that $\vert\langle\Psi\vert L_{A,ij}\otimes L_{B,mn}\vert\Psi^{\ast}\rangle\vert =
\vert\langle\chi_{ij,mn}\vert\Psi\rangle\vert\Psi\rangle\vert\,.$
So from Eq. (\ref{5}), we have
\begin{align}
ALB_{ij,mn}(\rho)&=\min_{\lbrace p_{k},\vert\Psi_{k}\rangle\rbrace} \sum_{k}p_{k}\vert\langle\chi_{ij,mn} \vert\Psi_{k}\rangle\vert\Psi_{k}\rangle\vert \,\notag \\
&=\min_{\lbrace p_{k},\vert\Psi_{k}\rangle\rbrace} \sum_{k}p_{k}\vert\langle\Psi_{k}\vert L_{A,ij}\otimes L_{B,mn}\vert\Psi_{k}^{\ast}\rangle\vert
\label{55}
\end{align}
 From Eq. (\ref{7}) and the above equation, we deduced that 
 \begin{align}
 C_{ij,mn}(\rho)= ALB_{ij,mn}(\rho) 
 \label{555}
\end{align}
 and so:
\begin{align}
\tau(\rho)=\sum_{i<j,m<n} ALB_{ij,mn}^{2}(\rho).
\label{8}
\end{align} 

Therefore, from the Eq. (\ref{17}) and Eq. (\ref{555}) we deduce that the Eq. (12) of Ref.~\cite{3}, i.e.
\begin{align}
C^{2}_{ij,mn}(({\bf{1}}\otimes\mathcal
{S})|\Psi\rangle\langle\Psi|)\,\qquad\qquad\qquad\qquad\qquad\qquad\notag \\
=\frac{d^{2}}{4}\big(\sum_{l>k=0}^{d-1}C_{ij,kl}(\vert\Psi\rangle)C_{kl,mn}(({\bf{1}}\otimes\mathcal{S})|\phi^{+}\rangle\langle\phi^{+}|)\big)^{2}.
\label{155} 
\end{align}
and so the Eq. (15) of the same reference, i.e.
\begin{equation}
\tau(({\bf{1}}\otimes\mathcal
{S})|\Psi\rangle\langle\Psi|)\geq\frac{2d\eta}{d-1}\frac{d^{2}}{4}\tau(({\bf{1}}\otimes\mathcal
{S})|\phi^{+}\rangle\langle\phi^{+}|)C^{2}(|\Psi\rangle).
\label{133}
\end{equation}
where $\eta=\min_{\{p,r\}}\omega_{p}\omega_{r}$ for any pair $p<r$
satisfying $\omega_{p}\omega_{r}\neq0$, dose not hold in general. 

%%%%%%%%%%%%%%%%%%%%%%%%%%%%%%%%%%%%%%%%%%%%%%%%%%%
\section{example}
Consider a two-qutrit system which one of its qutrit interacts with an environment. The time evolution of this 
system is given by the following Master equation:
\begin{align}
\dot\rho={\cal L}\rho\,,\qquad
{\cal L}=1_{A}\otimes {\cal L}_{B}\,,
\label{38}
\end{align}
\begin{center}
\begin{figure}[ptb]
\epsfig{file=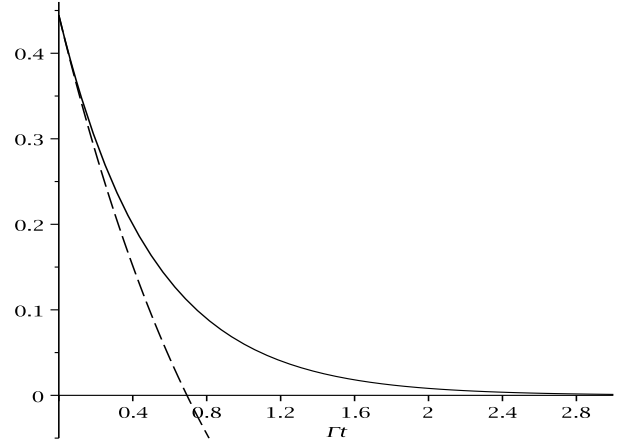,width=8.5 cm,height=6 cm}
\caption{Time evolution of the $MLB^{2}_{(1)12,12}$ when the initial state of the system is
$\vert\phi^{+}\rangle =\frac{1}{\sqrt{3}}(\vert 00\rangle +\vert 11\rangle+\vert 22\rangle)$ for the cases(a) spontaneous decay(dashed line)(b) decoherence(solid line).}
\label{fig1}
\end{figure}
\end{center}
where ${\cal L}_{B}$, for a one-qutrit $\rho_{B}$, is 
\begin{equation*}
{\cal L}_{B}=\dfrac{\Gamma}{2}\left(2\gamma\rho_{B}\gamma^{\dagger}-\rho_{B}\gamma^{\dagger}\gamma -
\gamma^{\dagger}\gamma\rho_{B}\right)\,. 
\end{equation*}
$\Gamma$ is the decay constant and $\gamma$ is a coupling operator characterizing the dynamics of system. 
For $\gamma=\left(
\begin{array}{ccc}
0 & 0 & 0  \\
\sqrt{2} & 0 & 0 \\
0 & 1 & 0 
\end{array}
\right)\,   \label{reduced}$ the Eq. (\ref{38}) represents the spontaneous decay of the system and for 
$\gamma=\left(
\begin{array}{ccc}
2 & 0 & 0  \\
0 & 1 & 0 \\
0 & 0 & 0 
\end{array}
\right)\,   \label{reduced}$ the Eq. (\ref{38}) represents the system's decoherence~\cite{12}.
 
In order to evaluate the entanglement dynamics of this system, we use the $MLB^{2}_{(1)ij,mn}(\rho)$ (which is a lower bound of squared concurrence). Fig. 1 shows the time evolution of $MLB^{2}_{(1)ij,mn}(\rho)$ for the case $i=1$, $j=2$, $m=1$ and $n=2$, when the initial state of the system is
$\vert\phi^{+}\rangle =\frac{1}{\sqrt{3}}(\vert 00\rangle +\vert 11\rangle+\vert 22\rangle)$(for other value of i,j,m and n, $MLB^{2}_{(1)ij,mn}(\rho)$ dose not give better estimate for entanglement).      
From this figure and using Eq. (\ref{15}), we can deduce the behavior of $MLB^{2}_{(1)12,12}(\rho)$ 
 for any initial states of the form $\vert\psi\rangle =a\vert 00\rangle +b\vert 11\rangle+c\vert 22\rangle$. 
For any such initial state, the ability of the $MLB_{(1)12,12}$ in detecting the entanglement of
 $\rho'=({\bf{1}}\otimes\mathcal {S})|\Psi\rangle\langle\Psi|$ is determined by the ability of $MLB_{(1)12,12}$ in detecting the entanglement of $\rho_{\mathcal {S}}=({\bf{1}}\otimes\mathcal {S})|\phi^{+}\rangle\langle\phi^{+}|$, which is shown in Fig. 1. Also, the amount of the lower bound $MLB_{(1)12,12}(\rho')$ is, up to a factor, equal to $MLB_{(1)12,12}(\rho_{\mathcal {S}})$.

%%%%%%%%%%%%%%%%%%%%%%%%%%%%%%%%%%%%%%%%%%%%%%%%%%%%%%%%%%%%%%%%%%%%%%%%%%%%%%%%%%%%%%%%

\section{conclusions}
We have studied the dynamics of two lower bounds of bipartite concurrence introduced in Eq. (\ref{5}) and 
Eq. (\ref{6}), when one party goes through an arbitrary channel. In Eq. (\ref{15}), we have shown that for arbitrary bipartite quantum states, $ MLB_{(1)ij,mn}(\rho)$ obeys the factorization law similar to that of Eq. (\ref{10}) for the concurrence. In an example, we have discussed the application of this factorization law in determining the behavior of the $ MLB_{(1)ij,mn}(\rho)$ in estimating the entanglement of the system. Also, we have shown that the $ ALB_{ij,mn}(\rho)$ obeys a similar factorization law for concurrence as Eq. (\ref{11}).

%%%%%%%%%%%%%%%%%%%%%%%%%%%%%%%%%%%%%%%%%%%%%%%%%%

\section*{acknowledgments}
The authors would like to thank Safa Jami for useful discussions. 

%%%%%%%%%%%%%%%%%%%%%%%%%%%%%%%%%%%%%%%%%%%%%%%%%%%

\end{document}